\documentclass{eas}
\usepackage{graphicx}
\newcommand{\Msun}{\mbox{\,$M_\odot$}}

\newcommand{\mic}{\mbox{$\,\mu$m}} 

\newcommand{\sirtf}{\mbox{\it Spitzer Space Telescope}}

\TitreGlobal{The Physics of Evolved Stars}
\begin{document}

\title{Infrared Observations of Novae in the SOFIA Era} 
\author{R. D. Gehrz}\address{Minnesota Institute for Astrophysics, 
School of Physics and Astronomy, 116 Church Street, S.E., University of Minnesota, Minneapolis, Minnesota 55455, USA}
\author{A. Evans}\address{Astrophysics Group, Keele University, Keele, Staffordshire, ST5 5BG, UK}
\author{L. A. Helton}\address{SOFIA Science Center, USRA, NASA Ames Research Center, M.S. 232-12, Moffett Field, CA 94035, USA}
\author{C. E. Woodward}\sameaddress{1}
\vspace{-4mm}
\begin{abstract}
Classical novae inject chemically enriched gas and dust into the local inter-stellar medium (ISM). Abundances in the ejecta can be deduced from infrared (IR) forbidden line emission.  IR spectroscopy can determine the mineralogy of grains that grow in nova ejecta.  We  anticipate the impact that NASA's new Stratospheric Observatory for Infrared Astronomy (SOFIA) will have on future IR studies of novae.
\end{abstract}
\maketitle
\vspace{-4mm}
\section{Introduction}
Classical novae (CNe), thermonuclear runaways (TNRs) that occur on the surfaces of white dwarfs (WDs) that are accreting matter in close binary systems, participate in the cycle of Galactic chemical evolution by ejecting gas and dust into the interstellar medium (ISM) where it is incorporated into new generations of stars and planetary systems.  We describe here how infrared (IR) observations, especially those expected to be conducted with (SOFIA), can be used to study grain composition and elemental abundances in nova ejecta.
\vspace{-2mm}
\section{Temporal Development of Classical Novae}
The physics of a CN eruption, its subsequent IR development, and its potential for enrichment of the ISM  have been reviewed extensively by Gehrz et al. (\cite{ge98}), Bode \& Evans (\cite{be08}), Evans \& Gehrz (\cite{eg12}), and Gehrz et al. (\cite{ge14}). The TNR causes the ejection of a discrete expanding shell that is enriched in metals by both the TNR and material scoured from the WD. At first, the ejected material appears as an optically thick fireball that radiates like a blackbody.  As the ejecta expand and disperse,  the IR spectrum is dominated by an optically thin free-free continuum overlain by strong hydrogen re-combination lines.  At later times, when the shell density falls below n$_{H}$ $\sim$ $10^{6}$ - $10^{7}$ cm$^{-3}$, the typical critical density for quenching forbidden lines by electron de-excitation, strong IR metallic forbidden lines dominate the spectrum in CNe that result from TNRs on ONe WDs. The prototypical ONe nova, QU Vul (1984\#2), showed strong forbidden line emission that persisted for more than twenty years (Gehrz et al. \cite{ge08}).  In some TNRs on CO WDs, the optically thin gas phase is followed by a dust formation event. Grains begin to nucleate in the expanding ejecta - when the radiation temperature from the central engine falls to $\simeq$ 1000K - and grow very rapidly thereafter (Shore \& Gehrz \cite{sh04}). The most extreme cases of dust production, typified by NQ Vul (Ney \& Hatfield \cite{ne78}) and LW Ser (Gehrz et a. \cite{ge80}) cause a pronounced visual extinction event as dust completely absorbs the luminosity of the central engine and re-radiates it in the IR. Grain types that have been observed to form in the ejecta of CNe include amorphous carbon, SiC, silicates, and hydrocarbons.  In at least two cases, all four grain types have formed at different times in the ejecta of a single nova. Metal abundances in the ejecta can be deduced from both IR dust emission features and forbidden line emission. Some ONe CNe have produced ejecta that are extremely overabundant in C, N, O, Ne, Mg, Al, and Si (Gehrz et al. \cite{ge98}).  The most extreme case is QU Vul, where neon was $\geq$ 168 times solar abundance (Gehrz et al. \cite{ge08}).  Over abundances of 20 to 75 of C, N, and O have been reported for a number of other novae that have been observed spectroscopically in the IR.  It seems probable that CNe can contribute significantly to abundance anomalies in neaby star-forming clouds.  In particular, they are theoretically capable of producing large amounts of the radio-isotope $^{22}$Na (half-life 2.7 days) that may be responsible for the over abundance of $^{22}$Ne in solar system meteorites.
\begin{table}[htb!]
\caption{Selected IR Forbidden Lines within FORCAST Grism Passbands.}
\includegraphics[width=12.5cm]{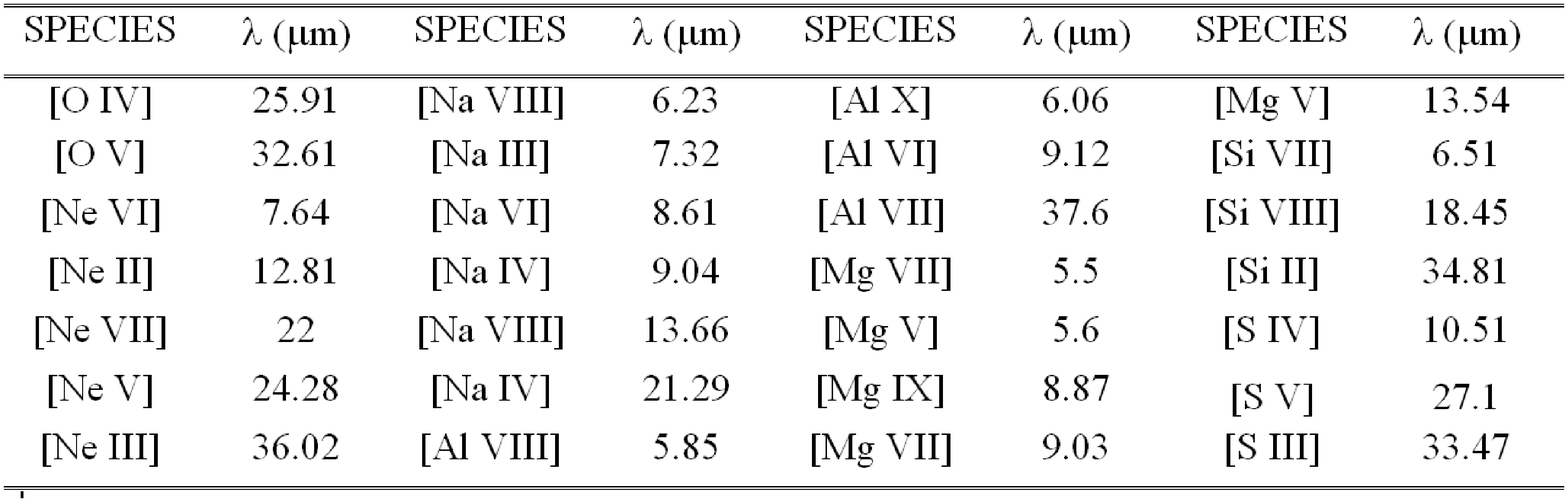}
\end{table}
\vspace{-2mm}
\section{SOFIA as a Platform for IR Observations of CNe}
SOFIA (Figure 1, Young \etal\ \cite{yo12}), a joint project of NASA and the German Space Administration (DLR), is a 2.5-meter infrared airborne telescope in a Boeing 747-SP flying in the stratosphere at altitudes as high as 45,000 feet (13.72 km) where the atmospheric transmission averages more than 80 \%  throughout the 0.3 - 1600\mic\ spectral region. Two SOFIA science instruments, The First Light Infrared TEst CAMera (FLITECAM, McLean et al. \cite{mc06}) and The Faint Object infraRed CAmera for the SOFIA Telescope (FORCAST, Herter et al.\cite{he12}), have grisms covering the entire 1 to 40\mic\ spectral region required for studying dust grain mineralogy and IR forbidden line emission with the required spectral resolution (R = $\lambda/\Delta\lambda$ of $\approx$ 100-200). FORCAST can measure the strengths of many of these lines, a number of which are unobservable from the ground because of atmospheric absorption (see Table 1). Even though CNe are predicted to produce large amounts of $^{22}$Na, none of the Na lines shown in Table 1 have yet been detected in a CN.  SOFIA may provide the best chance yet for detecting these lines. The FORCAST grisms will also enable observations of the spectral structure associated with crystalline silicate grains and the unidentified IR (UIR) features believed to be associated with polycyclic aromatic hydrocarbons (PAHs). Additional forbidden lines and PAH features lie within the wavelength regions covered by the FLITECAM grisms.  SOFIA has unique potential for greatly expanding our understanding of the IR temporal development of CNe. It can fly anywhere and any time to enable observations of targets of opportunity and can observe objects as close as 20 degree to the Sun.  Cryogenic IR space platforms like \sirtf\ and {\it James Webb Space Telescope (JWST)} have very limited viewing windows because they are constrained to look away from the Sun.  Furthermore, SOFIA can see more IR forbidden lines than any existing or planned mission except {\it JWST}. The SOFIA Program Announces Observing Opportunities on an annual basis.
\begin{figure}[htb!]
\includegraphics[width=5.75cm]{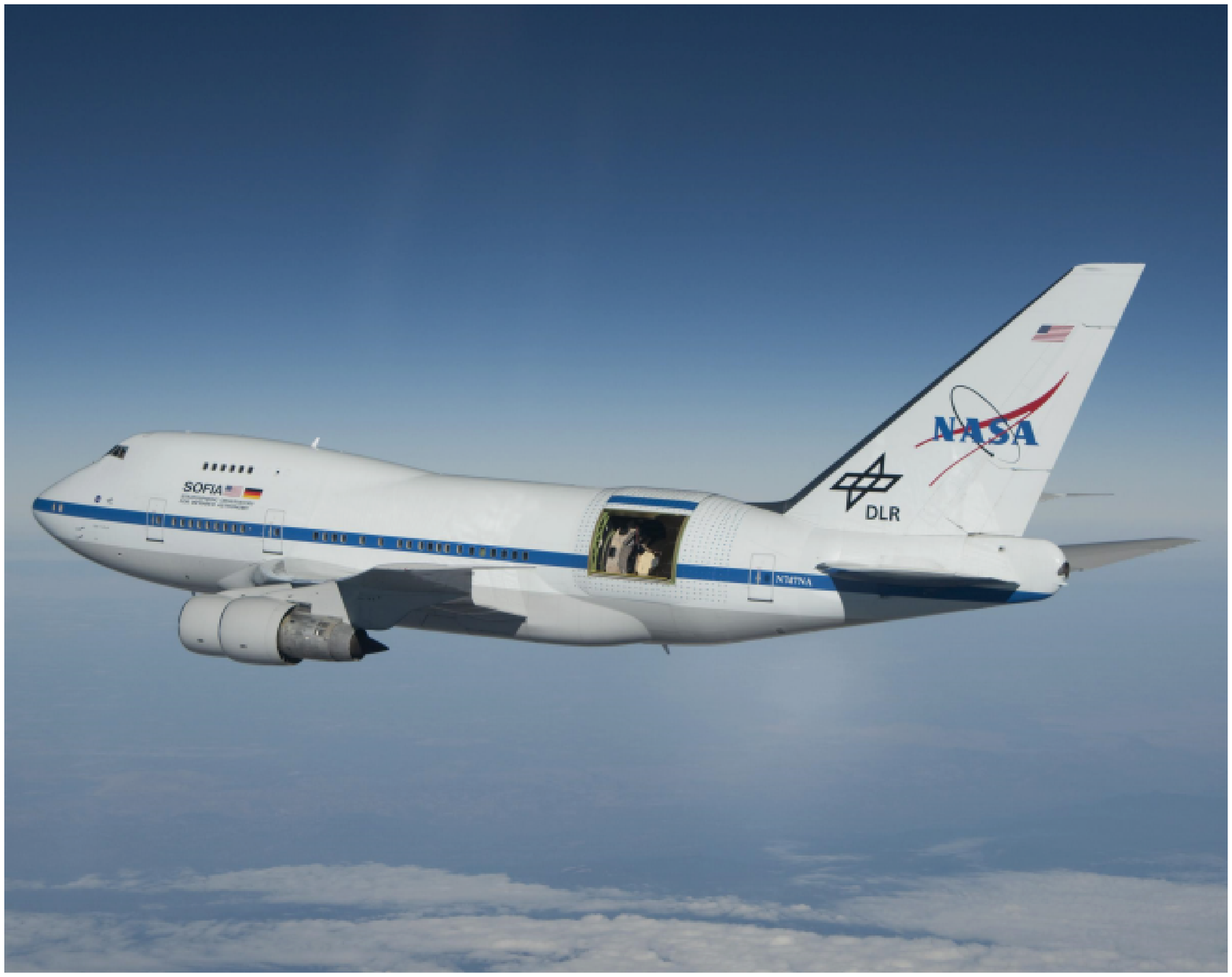}
\qquad
\includegraphics[width=6cm]{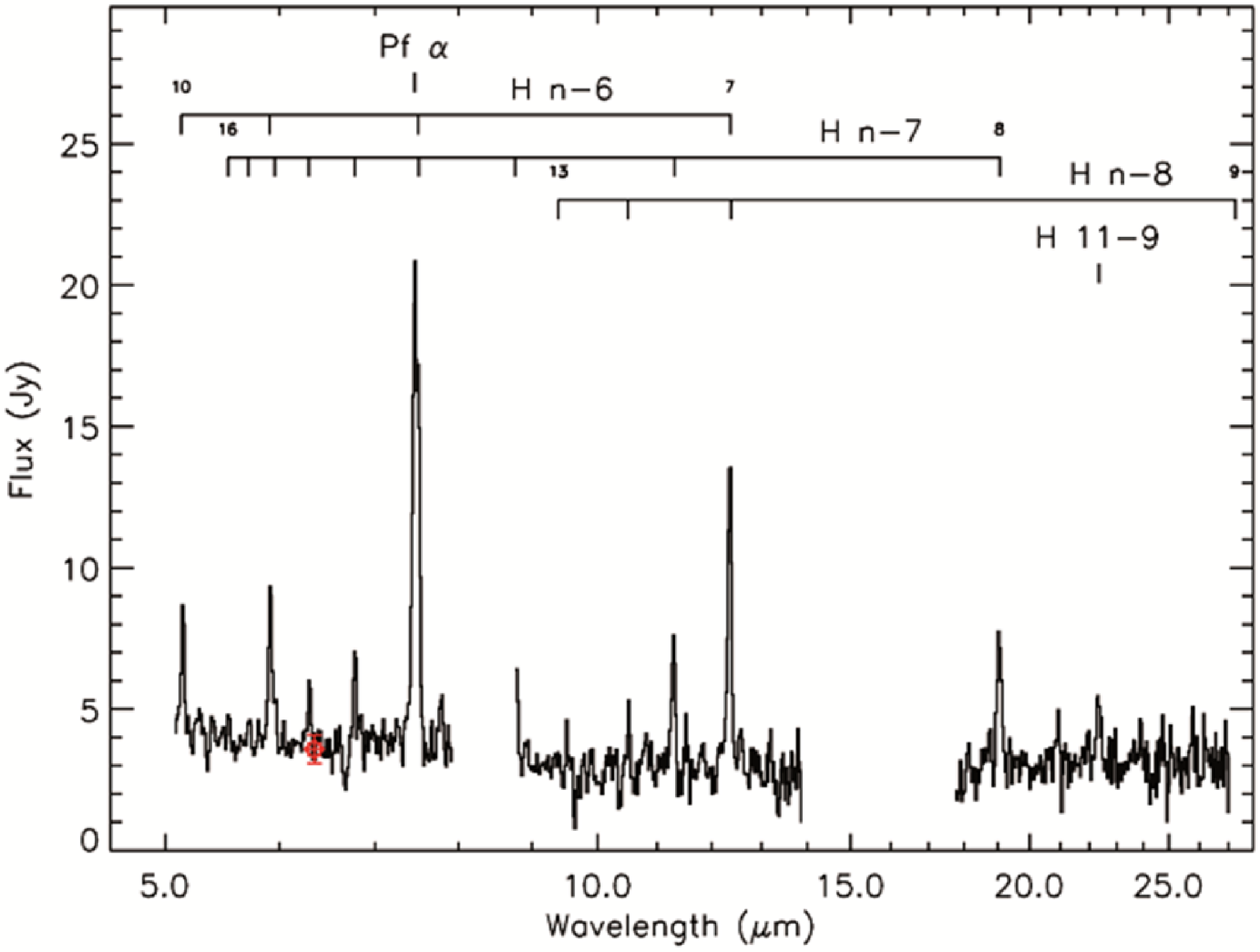}
\caption{{\bf Left:} SOFIA in flight. {\bf Right:} Hydrogen spectrum of V339 Del 27 days after outburst (Gehrz et al. \cite{ge15}).}
\end{figure}
\begin{figure}[htb!]
\includegraphics[width=12.5cm]{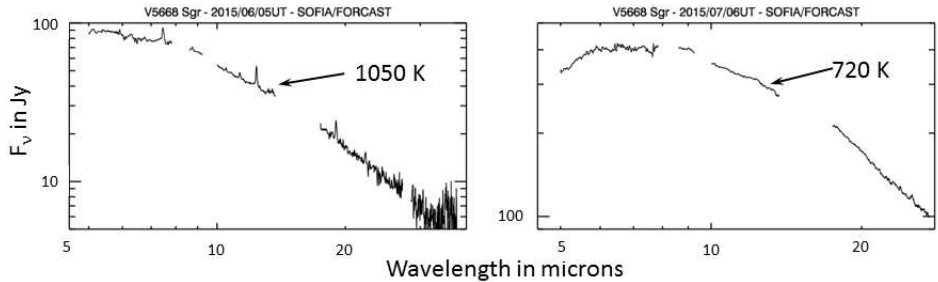}
\caption{SOFIA FORCAST spectra of dust formation in V5668 Sgr. Spectra {\bf Left:} Nucleation phase. {\bf Right:} Grain emission maximum.}
\end{figure}
Nova V339 Del (Nova Del 2013), the first CN to be observed with SOFIA, showed a rich hydrogen recombination line spectrum associated with the free-free development phase (Figure 1). The shell density was still n$_{e}$ $\geq$\ 10$^{11}$ cm$^{-3}$ so that both the forbidden lines and the hydrogen lines were quenched.  Nevertheless, we were able to use the SOFIA spectrum to estimate a lower limit to the ejected hydrogen gas mass of 5$\times$10$^{-6}$ - 10$^{-5}$ $\Msun$ (Gehrz et al. \cite{ge15}). We have followed the early phase of dust formation in Nova V5668 Sgr (2015\#2) with FORCAST (Figure 2).  A spectrum on 2015 June 05, that we interpret as showing the early nucleation phase, showed a $\sim$ 1100 K dust continuum overlain by weak hydrogen emission lines. On 2015 July 06, the dust emission had increased by a factor of $\sim$ 40, the grain temperature had fallen to $\sim$ 750 K, and the dust continuum had completely blanketed the hydrogen emission.  We attribute these effects to the rapid growth of the grains to sub-micron radii (see Shore \& Gehrz \cite{sh04}). 

\vspace{-4mm}
\section*{Discussion}
\vspace{-2mm}
\noindent {\bf Q: Orsola de Marco:} Do you think there could be a connection between classical novae and born again stars?\\
\mbox{} \hspace{3mm}  {\bf A: Bob Gehrz:} In principle, since the TNR depletes H and enriches He, the material left on the surface of the WD after the eruption has He that might provide a late thermal pulse under the right conditions. It is an interesting question for the theorists to ponder.\\

\end{document}